\documentclass[iop]{emulateapj}
\slugcomment{Accepted for publication in {\it The Astrophysical Journal} 9/09/2010}  
\def\lax {\ifmmode{_<\atop^{\sim}}\else{${_<\atop^{\sim}}$}\fi}  
\def\gax {\ifmmode{_>\atop^{\sim}}\else{${_>\atop^{\sim}}$}\fi}  
\def\gtorder{\mathrel{\raise.3ex\hbox{$>$}\mkern-14mu
             \lower0.6ex\hbox{$\sim$}}}

\begin{document}

\title{Discovery and Monitoring of a new  Black Hole Candidate  XTE J1752-223 with {\it RXTE}: 
RMS spectrum evolution, BH mass and the source distance.}

\author{Nikolai Shaposhnikov\altaffilmark{1,2}, Craig Markwardt\altaffilmark{2}, Jean Swank\altaffilmark{2} and Hans Krimm \altaffilmark{2,3}}

\altaffiltext{1}{CRESST/University of Maryland, Department of Astronomy, College Park MD, 20742, nikolai.v.shaposhnikov@nasa.gov}

\altaffiltext{2}{Goddard Space Flight Center, NASA, Astrophysics Science Division, Greenbelt MD 20771}

\altaffiltext{3}{CRESST/Universities Space Research Association, Columbia MD, 21044}

\begin{abstract}
We report on the discovery and monitoring observations of a new galactic black hole candidate
XTE J1752-223 by Rossi X-ray Timing Explorer ({\it RXTE}). The new source appeared on
the X-ray sky on October 21 2009 and was active for almost 8 months.
Phenomenologically, the source exhibited the low-hard/high-soft spectral state bi-modality and the 
variability evolution during the state transition that matches standard
behavior expected from a stellar mass black hole binary. We model the energy spectrum 
throughout the outburst  using a generic Comptonization model assuming that part of the 
input soft radiation in the form of a black body spectrum gets reprocessed in the Comptonizing medium. 
We follow the evolution of fractional root-mean-square (RMS) variability in the RXTE/PCA energy band
with the source spectral state and conclude that broad band variability is strongly correlated  with the 
source hardness (or Comptonized fraction). We follow changes in the energy distribution of rms 
variability during the low-hard state and the state transition and find further evidence that 
variable emission is strongly concentrated in the power-law spectral component. 
We discuss the implication of our results to the Comptonization regimes during different spectral states.
Correlations of spectral and variability properties provide measurements of the BH mass and distance to the source. 
The spectral-timing correlation scaling technique applied to the {\it RXTE} observations during
 the hard-to-soft state transition indicates  a mass of the BH in XTE J1752-223  between 8 and 11
solar masses and a distance to the source about 3.5 kiloparsec.
\end{abstract}


\section{Introduction}
A black hole (BH) X-ray transient is a binary system in which a stellar mass black hole is orbiting a regular
star. During mass exchange episodes that occur 
when a compact star oblates matter from the normal star a system is seen as a bright X-ray source. 
%onto the BH occurs through either Roche lobe overflow or a  strong stellar 
%wind. 
This  accretion process is accompanied by a plethora of spectral and variability phenomena
that are unique to BH candidates and most probably related to a strongly relativistic nature of the compact object. 
Among the most puzzling aspects of the accreting BH phenomenology is 
spectral state bi-modality,  extended power law spectral tails, aperiodic and quasi-periodic 
variability. Uncovering the nature of these effects is crucial to advance in different aspects of
BH physics.
%Uncovering the nature of these effects will allow better understanding 
%the physics of BH themselves. 
%which contain valuable information on these strongly relativistic objects.
%Properties and evolution of accreting black holes provide the most direct information on 
%the behavior of matter in the strong gravity limit. 
In this Paper we present {\it Rossi X-ray Timing Explorer (RXTE)} observations of a newly discovered BH candidate source XTE J1752-223 during its recent October 2009 - May 2010 outburst. 
We describe the general evolution of the outburst, correlation of the energy spectral characteristics and
variability properties and the changes in the variability distribution with energy. We estimate the BH mass
using the correlation scaling technique and address some implication of variability evolution
during the observed spectral state  transition.

The main feature of the X-ray spectrum observed from an accreting
BH is a presence of a strong non-thermal component described by a power law. 
The origin of this emission is attributed to multiple inverse Compton scattering of the soft photons 
off the energetic electrons near the central source. The properties and the balance between the 
thermal and non-thermal components in the source spectrum is the primary parameter
defining a BH spectral state \citep{rm,bell05}. The state with a dominant non-thermal emission is called
the low-hard state (LHS) due to the fact that it is usually observed during low luminosity
episodes during the rise and decay stages of an outburst. During most outbursts 
BH sources exhibit transitions (intermediate state, IS) to the high-soft state (HSS), dominated by 
the thermal component, attributed to a geometrically thin accretion disk.
Strong (up to 50 percent root-mean-square) aperiodic variability is seen in
Fourier Power Density Spectra (PDS) during the LHS in the form of band-limited noise
 which is flat at low frequencies  and a power-law at higher frequencies 
 (alternative names used in literature are flat top noise and white-red noise).
The rms fraction of this noise component tends to decrease 
in intermediate state and is suppressed to a few percent or not detected at all 
in the HSS. During the LHS and IS quasi-periodic oscillations (QPO) are also present in the PDS. The
hard/soft state dichotomy is one of the most puzzling aspects of the accreting stellar mass BH phenomenology. This behavior has a very recognizable pattern \citep[most commonly presented 
as a famous q-shaped hardness-intensity diagram, see][for the latest review]{bel10}. 
It is reproduced in almost all sources and outbursts with a striking level of stability. 

There are more than twenty X-ray sources in our galaxy in which optical measurements
show BH masses higher than the theoretical mass limit for a rotating neutron star \citep[see][ for details]{rm}. These sources
are therefore called confirmed BHs. There is also approximately the same number of X-ray binaries for which optical mass measurements are unavailable but which otherwise show properties very similar to the confirmed BHs. These sources are referred to as BH candidates. Better knowledge of the BH population is important for understanding some aspects of stellar evolution. In this Paper we present  BH mass and distance measurements for XTE J1752-223 using the correlation scaling method. \citet[][hereafter ST09]{st09} have shown that correlations during 
state transitions can be used to estimate BH mass and source
distance. The scaling method relies on the theoretically motivated and observationally tested 
assumption that the QPO frequency for a particular
accretion state of the system is set by the BH mass. Information on the 
spectrum normalization allows us also to estimate the distance to the source.
%The method is both teoretically motivated and observationally tested (see ST09 for details).
During our monitoring program with {\it RXTE} we were able to observe a part of the LHS-to-HSS 
transition, which provided sufficient data to apply the scaling method to XTE J1752-223.
The scaling method yields for XTE J1752-223 the BH mass of  $\sim 9.5$ M$_\odot$ and
the distance of about 3.5 kpc.

The main goal of this Paper is to present the evolution of X-ray properties 
during the XTE J1752-223 discovery outburst observed with {\it RXTE} and to report the mass and distance estimates of the compact  object based on  the  spectral and variability correlation scaling method.
We analyzed all RXTE observations taken during
the outburst active phase. We follow the properties of the energy spectrum by applying a Comptonization model
to the data. We also characterize the variability properties by examining  periodic and aperiodic 
features in the  Fourier Power Density Spectrum.
We also analyze the energy dependence of the variability and its evolution
during the LHS-to-HSS state transition. Specifically, we find
strong evidence that fast (less than 100 sec) 
aperiodic variability is almost completely confined to the power-law spectral component,
while the black body component introduces no significant variability.
The observed distribution of variability within the non-thermal part of the X-ray emission, i.e.
a steep decrease of variability amplitude with energy, has important implications for the physics
of photon up-scattering in the accreting BHs.

The paper is structured as follows.
In the next section we describe the discovery observation with {\it RXTE}/PCA bulge scans.
 Details of our data analysis are presented in \S \ref{data}.
 In \S \ref{evolution} we present the general evolution of the source 
as observed by with {\it RXTE} and {\it Swift},
and uncover the source evolution through BH spectral states. 
In \S \ref{rms_dist} we present analysis of the energy dependence of the
variability and its evolution with the source spectral states.
Mass measurements and distance estimates with {\it RXTE} data using the scaling technique 
are described in \S \ref{mass}. In \S \ref{discussion} we discuss the source behavior and 
possible implications of our observational results for the accretion regimes
in BHs during different spectral states and transitions between them. 
Conclusions follow in \S \ref{summary}.

\section{Discovery and identification of XTE J1752-223 as a new Galactic BH candidate}
\label{disc}

\begin{figure}
\includegraphics[scale=0.45,angle=0]{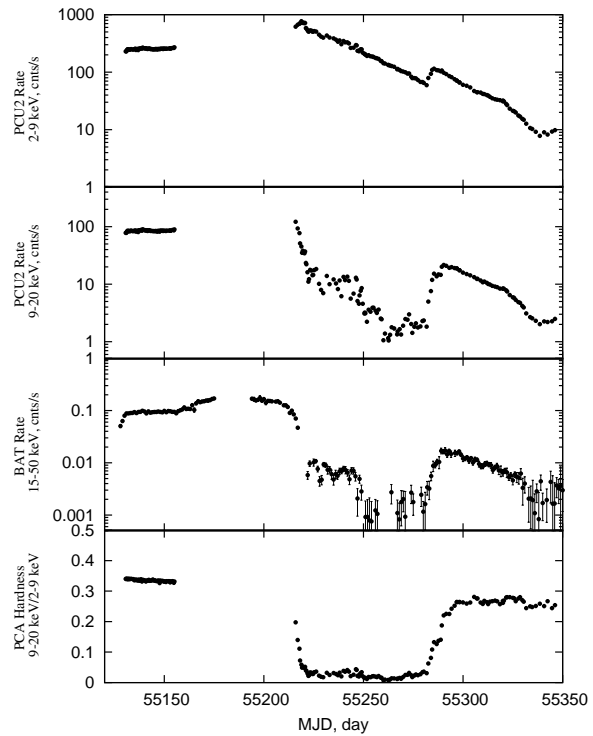}
\caption{{\it RXTE} and {\it Swift} lightcurves calculated for various  energy ranges. 
%periods of invisibility  for observation due to the Sun constrains for RXTE and Swift/BAT.
}
\label{curves}
\end{figure}

On October 23, 2009 17:52 UT  {\it RXTE} performed a routine scan of the Galactic bulge 
using the Proportional Counter Array (PCA). The scan analysis showed residuals consistent with
a new source at the position RA = 268.05$\pm$0.08, DEC = -22.31$\pm$0.02 (J2000 coordinate system).
Search for fast time variability yielded no significant pulsations. 
On October 24, 2009 {\it Swift} BAT instrument triggered on the source
and determined a position consistent with the PCA error circle \citep{atel2258}.
On  October 25, 06:08 UT {\it Swift} performed a dedicated follow-up observation, which
confirmed the previously determined  source position and allowed to estimate an interstellar extinction toward
the source of $N_H=0.46\times 10^{-22}$ cm$^{-2}$. The observation also revealed a very hard
non-thermal spectrum with a power law index of ~1.2 \citep{atel2261}.
On October 26, 15:03 UT  {\it RXTE} was able to make its first pointed observation of XTE J1752-223.
Overall spectral and timing properties are strongly reminiscent of the extreme low-hard states exhibited by 
Cyg X-1 and GX 339-4, which prompted a preliminary classification of the source as a BH candidate \citep{atel2269}.
Following this identification we triggered our {\it RXTE} Cycle 14 TOO observations for frequent monitoring of
BH transient X-ray sources. This program revealed source evolution completely consistent
with the source being a BH candidate. Namely, the source exhibited typical black hole spectral states,
accompanied with corresponding fast variability properties.
As described below, this monitoring completely confirmed our expectations
and showed source evolution and state transitions consistent with the current
observational scenario of classical accreting BH in outburst. Specifically, XTE J1752-223 
went through the initial LHS, transitioned through IS to HSS, returned to LHS and decayed to quiescence. 
The very high state, sometimes called the steep power law state \citep{rm}, which occurs for the most 
luminous BH transients, was not observed. 
The nature of the central compact object in XTE J1752-223 as a BH is also strongly supported by
 BH mass estimate presented in \S \ref{mass}, which strongly favors the mass range of 8-11 M$_\odot$.

\section{  {\it RXTE} Observations and Data Analysis}
\label{data}

We have analyzed 195 pointed {\it RXTE}/PCA observations beginning on October 26 2009, 14:24 UT (MJD 55130.6) 
and ending on April 20, 2010 22:48 UT (MJD 55306.95).
For the period of two months beginning from November, 20 2009 until January 19,
2010 XTE J1752-223 was not available for RXTE pointed observations due to the Sun 
constraint. {\it Swift}/BAT also was not able to observe the source for approximately a month within  this
period. While the source showed some activity after April 20, 2010, 
we do not include {\it RXTE} observations after this date in our analysis
as the PCA data is dominated by the Galactic ridge emission.  
For spectral analysis we utilized data collected with the PCU 2 top layer only.
We extracted spectra from Standard2 mode data files which provide spectra in 129 channels. The PCA response 
was calculated using the FTOOL {\it pcarmf 11.7}. Spectra were deadtime 
corrected according to "RXTE Cookbook"\footnote{http://heasarc.gsfc.nasa.gov/docs/xte/recipes/cook\_book.html}. 
The PCA background was estimated using FTOOL task
{\it pcabackest}. For spectral analysis the background was scaled to match the detected 
counts in Standard2 channels above 110 corresponding to energies above 75 keV,
where no significant source signal is expected. 

We extracted the background subtracted {\it RXTE}/PCA count rates 
in energy bands 2-9, 9-20 and 20-50 keV. 
{\it RXTE} and {\it Swift}/BAT lightcurves along with the hardness ratio are presented in Figure \ref{curves}.  
The {\it Swift}/BAT data is provided by the {\it Swift}/BAT team.

\begin{figure}
\includegraphics[scale=0.45,angle=0]{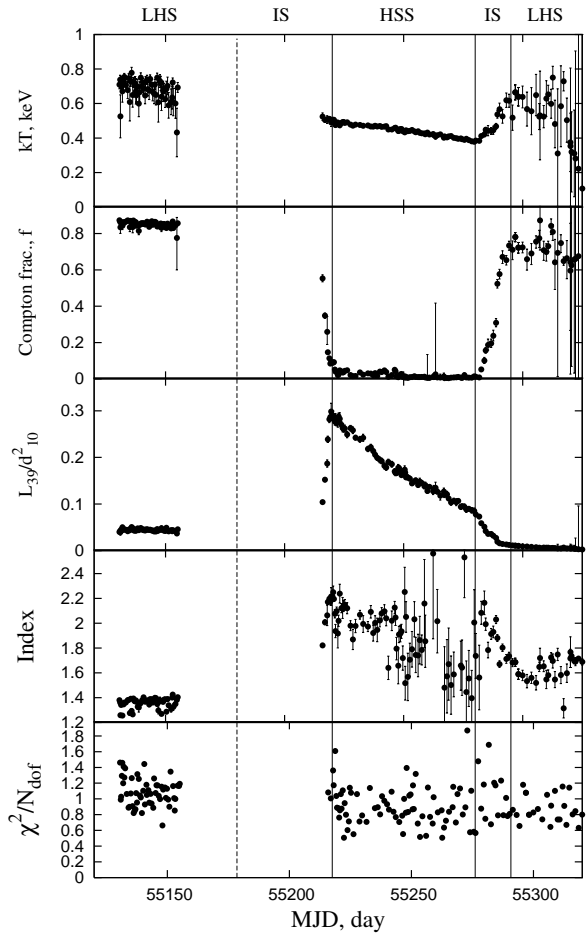}
\caption{ Evolution of XTE J1752-223 spectral parameters throughout the outburst inferred for individual {\it RXTE}/PCA poiting 
observations. The labels at the top show the corresponding spectral state of the source. The boundary between initial the LHS and the IS is shown approximately as no information is available due to the Sun occultation.}
\label{fits}
\end{figure}

\begin{figure*}
\hspace{0.1in}\includegraphics[scale=0.17,angle=-90,clip=false]{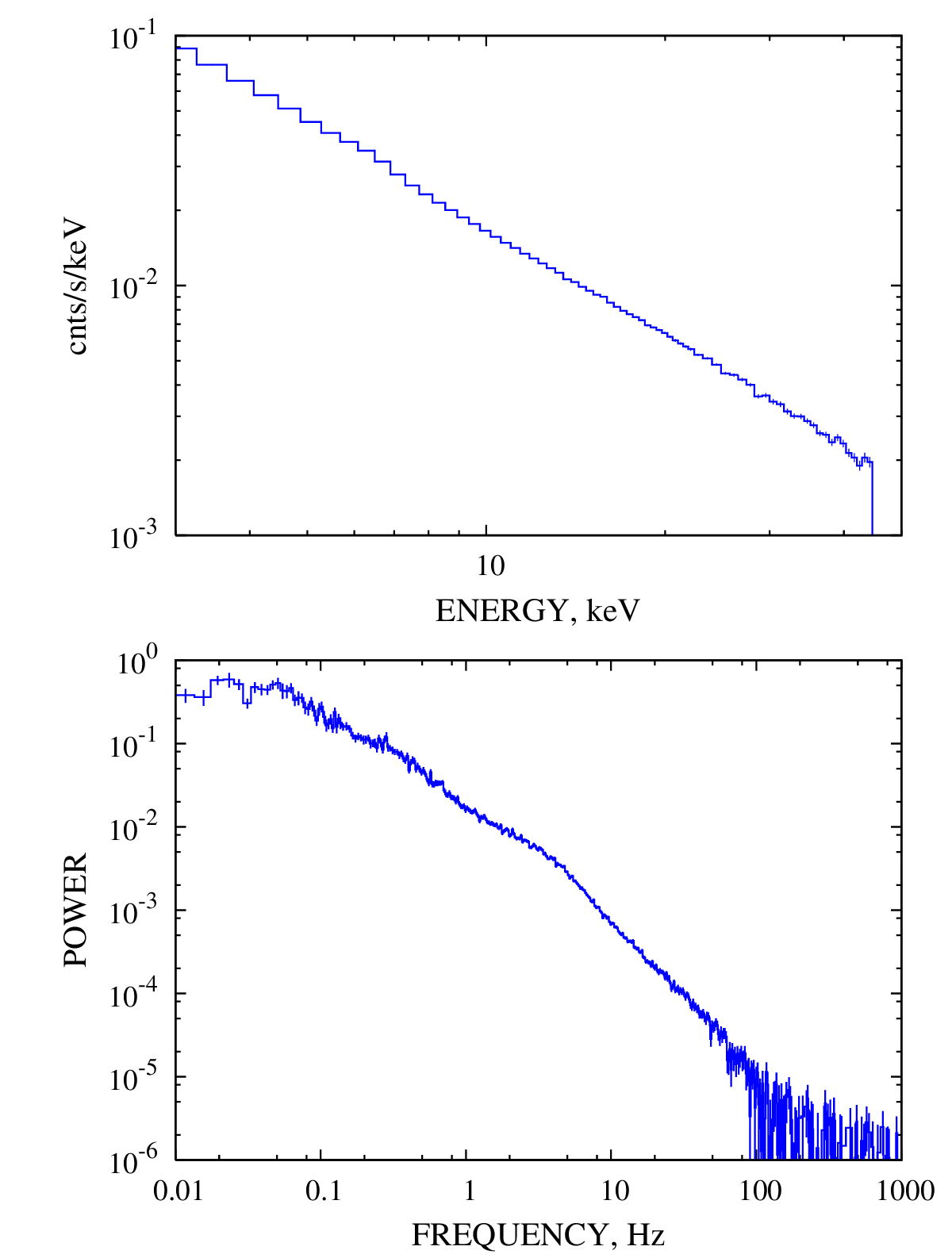}
\hspace{-0.17in}\includegraphics[scale=0.17,angle=-90,clip=true]{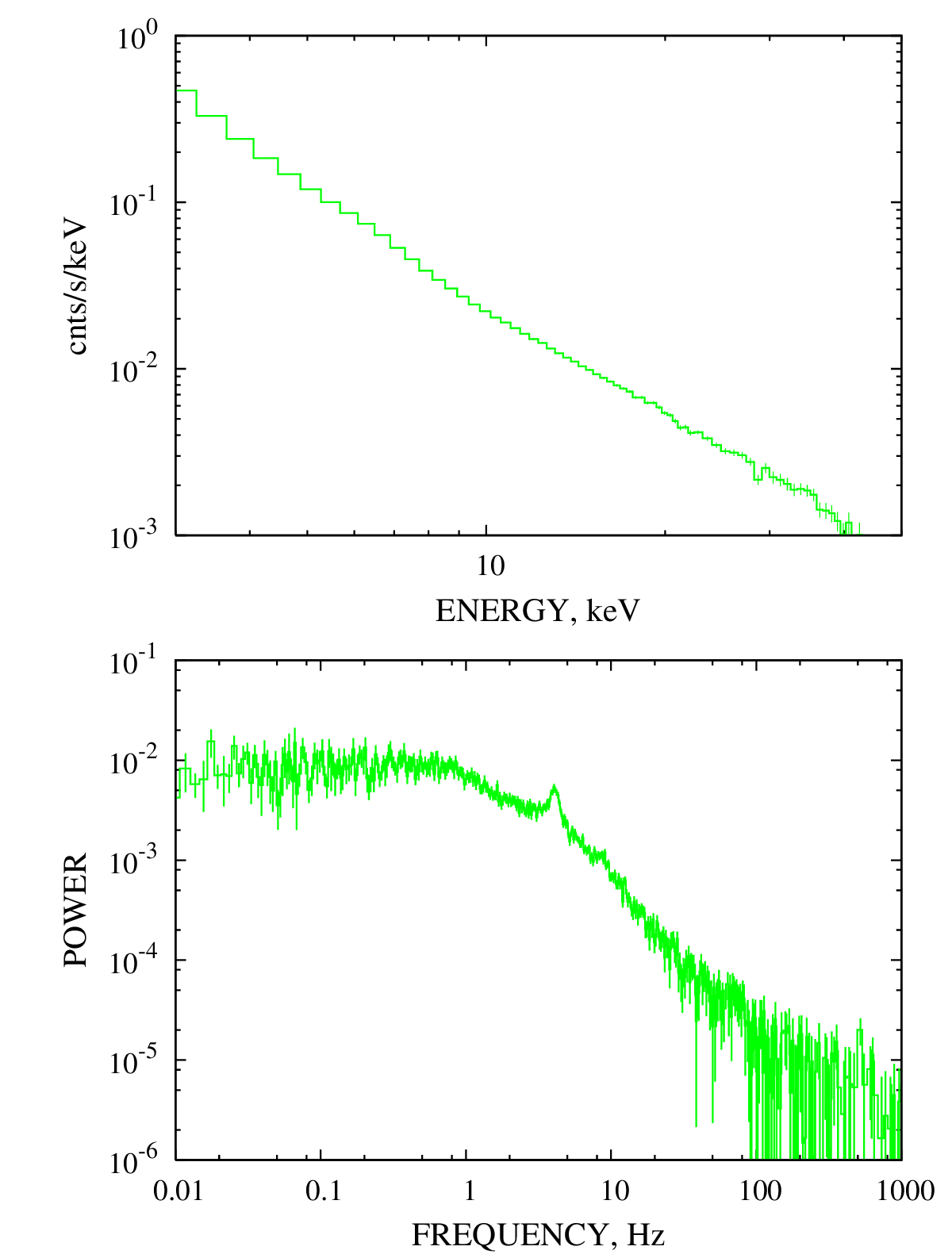}
\hspace{-0.17in}\includegraphics[scale=0.17,angle=-90,clip=true]{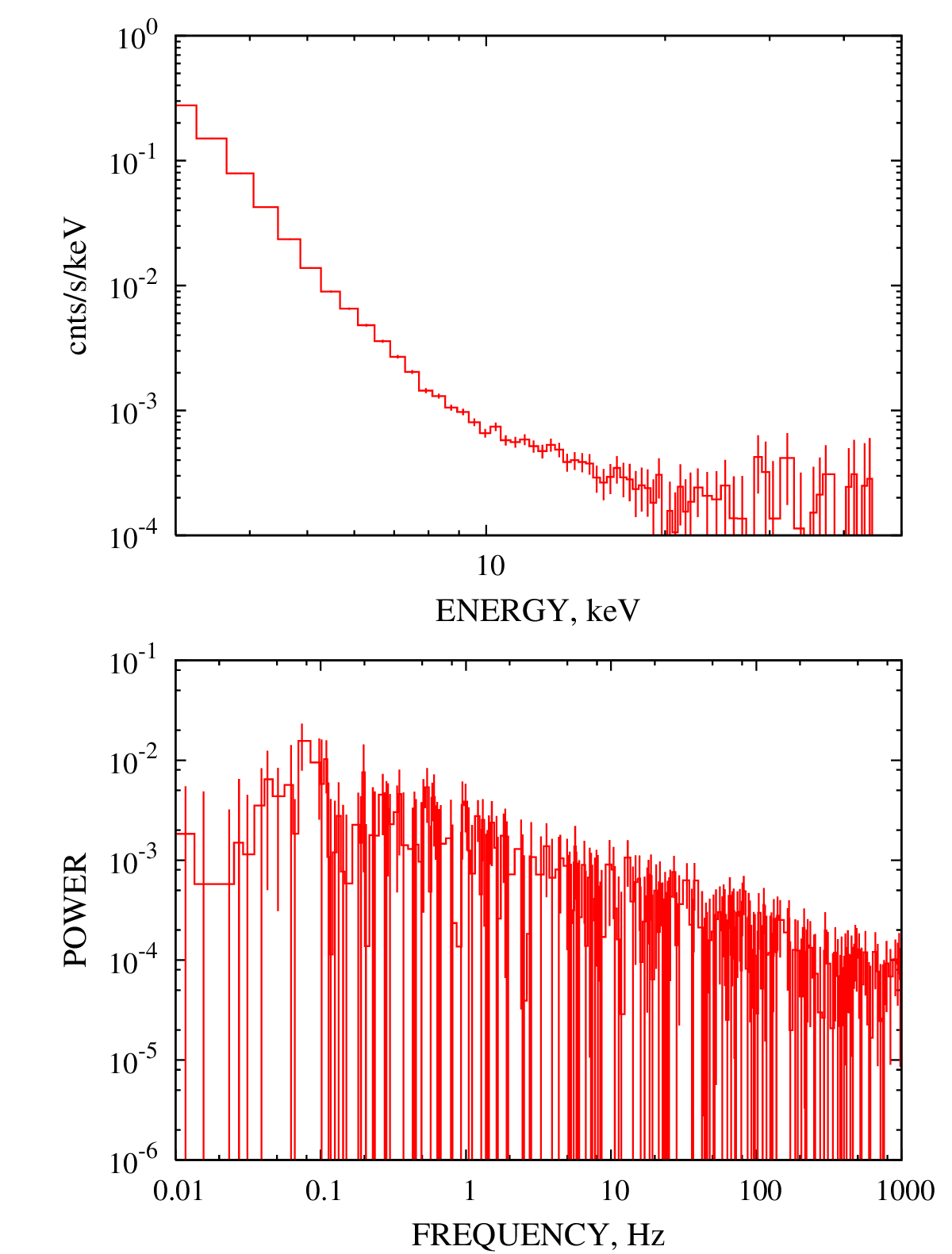}
\hspace{-0.17in}\includegraphics[scale=0.17,angle=-90,clip=true]{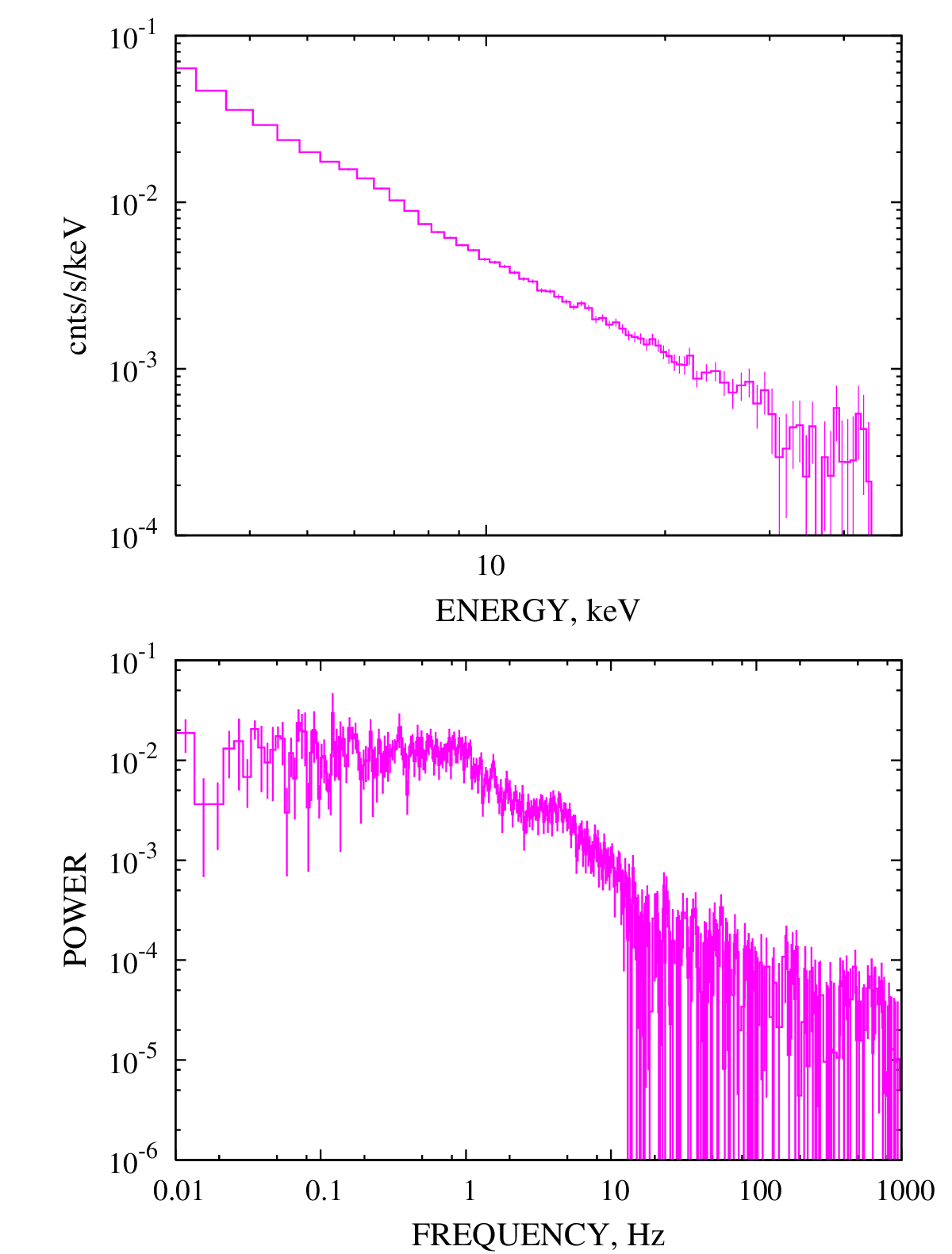}
\hspace{-0.17in}\includegraphics[scale=0.17,angle=-90,clip=true]{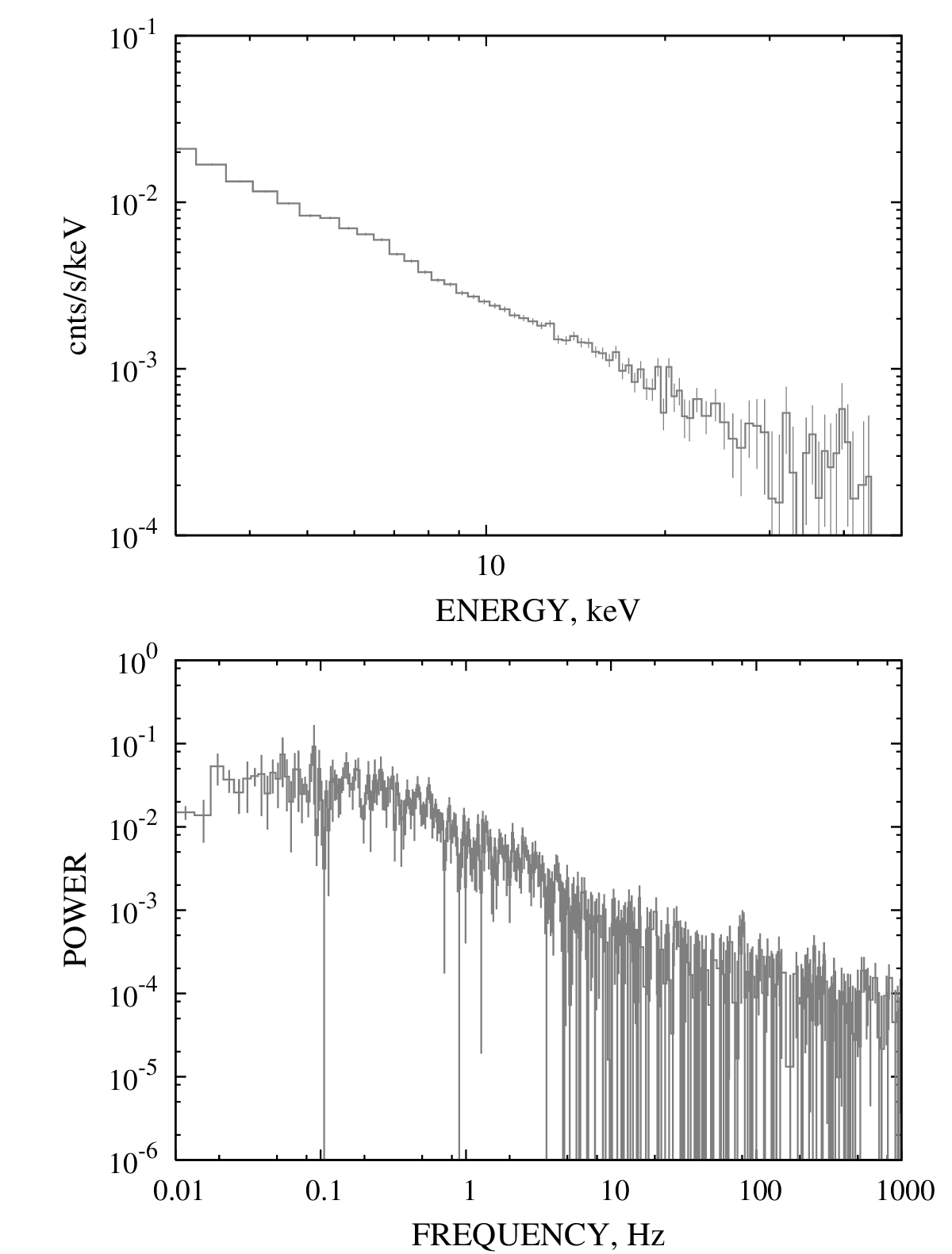}

\includegraphics[scale=0.213,angle=-90,clip=true]{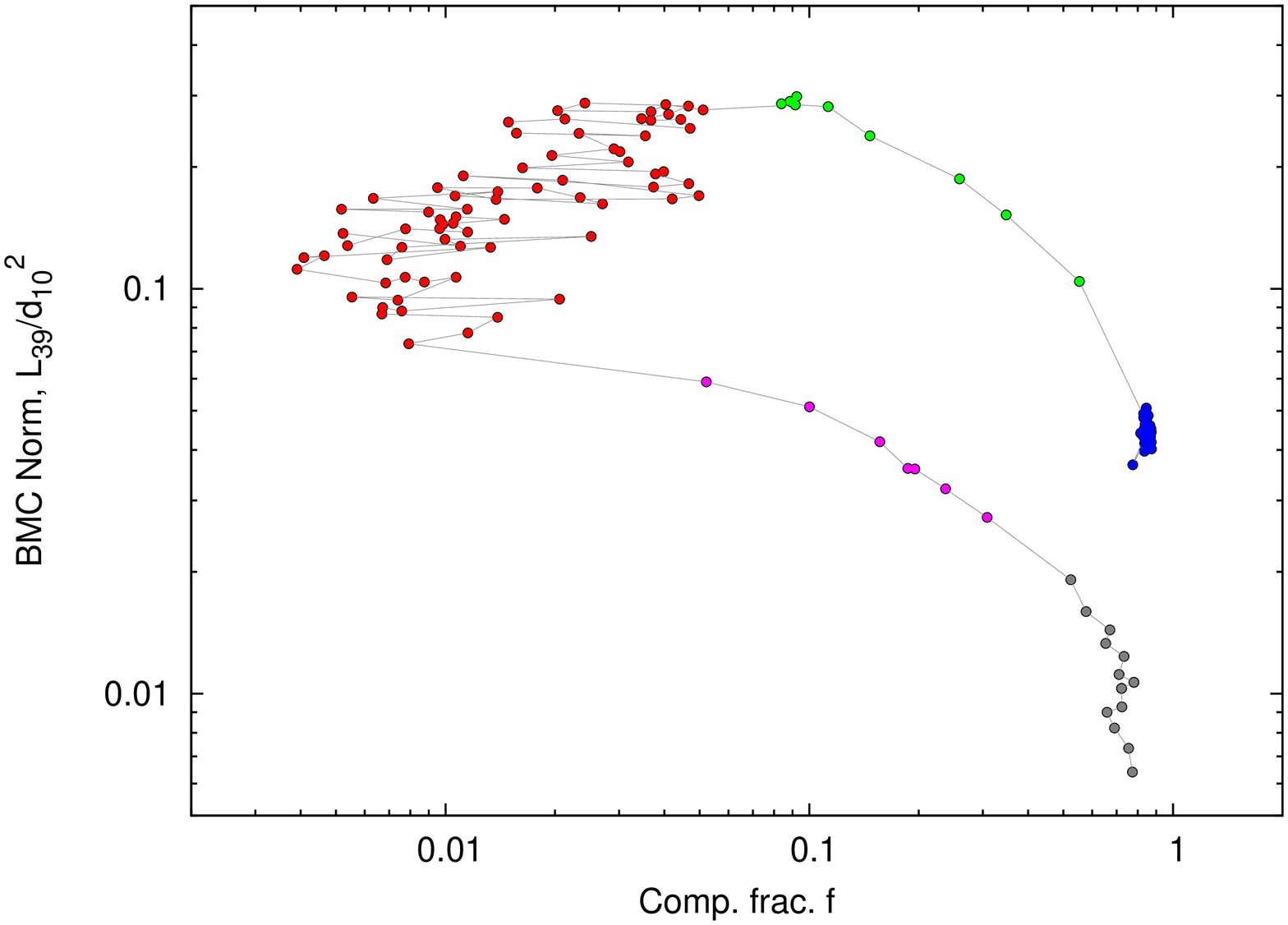}
\includegraphics[scale=0.213,angle=-90,clip=true]{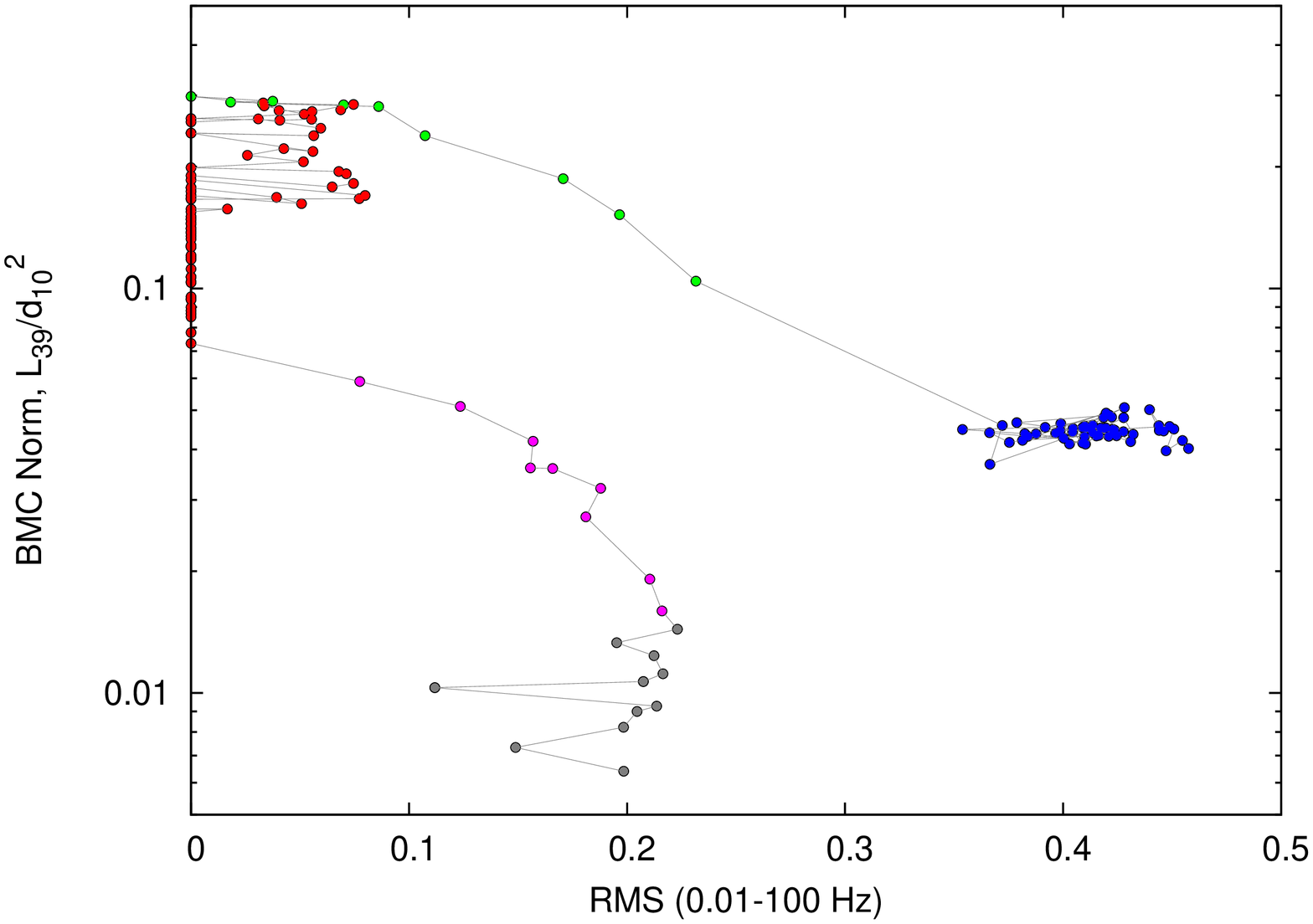}
\includegraphics[scale=0.213,angle=-90,clip=true]{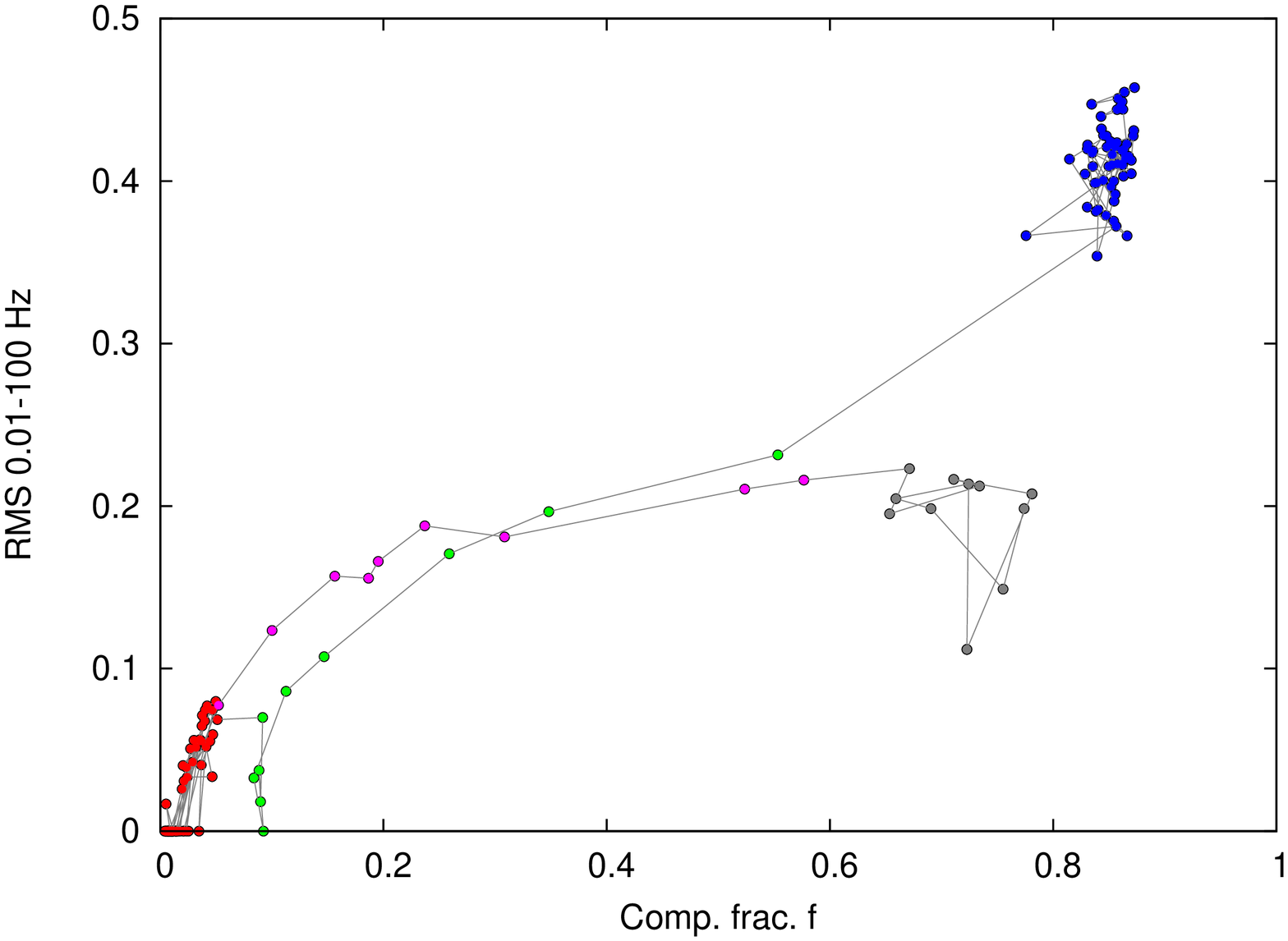}

\caption{Correlation of spectra and variability in XTE J1752-223 during the reported outburst. 
Representative energy spectra are shown in the top panels, while the power density spectra are
presented in the second row.  In the bottom row the hardness-intensity,  hardness-rms and rms-flux 
correlations are shown. The colors correspond to spectral states as follows: blue for the initial LHS, green for the rise
IS, red for the HSS, magenta for the decay IS and grey for the decay LHS. The specific RXTE observation used for the energy and power spectra are: 94044-07-01-000 for  the initial LHS, 94331-01-06-01 for the rise
IS, 95360-01-05-04 for the HSS, 95360-01-11-01 for the decay IS and 95702-01-01-00 for the decay LHS. }
\label{corr}
\end{figure*}

PCA spectra were analyzed in the energy range from 3.0 to 45.0 keV. 
To model the spectra we apply the model consisting of the generic Comptonization model 
\citep[BMC XSPEC model;][]{bmc} plus a {\it gaussian} for the iron emission line at 6.4 keV, modified 
by the interstellar absorption with $N_H$ fixed at $0.46\times 10^{22}$ cm$^{-2}$ \citep{atel2261}.
The main component of our spectral model was initially introduced to treat spectra
produced by boosting thermal disk photons in the convergent inflows near the central BH
where bulk motion is dominant in upscattering, hence the name BMC (Bulk Motion Comptonization).
The model employs the Green function convolution technique to convert a
part of the input black-body spectrum to the power law component. Designed in this
way the BMC model is applicable to any upscattering process which can be described 
by a broken power-law Green function and is applicable to the Comptonization on thermal
electrons. The parameters of the BMC model are the temperature of the input black body
spectrum $kT$, its normalization $N$ in units of $(L/10^{39} {\tt erg/s})(10 {\tt kpc}/d)^2$, the Comptonized fraction $f$ (implemented as a $log(A)$ parameter, where $f=A/(1+A)$) and the photon index $\alpha=\Gamma-1$ of the power law tail.
The BMC model  does not account for the electron recoil during Comptonization as the recoil 
effect  strongly depends on a particular Comptonization regime and is not trivial
to treat analytically. It therefore has to be modeled phenomenologically 
by adding the high energy cutoff to X-ray spectra.
During the LHS the high energy cutoff above $\sim$20 keV
was added to the model as required by the data. Evolution of spectral parameters of the Comptonization model are
given in Figure \ref{fits}. The analysis was done in XSPEC 12.0 analysis package. 

For timing analysis we constructed  Fourier power density spectra (PDS) for each 
observation in the frequency range from 0.01 Hz to 512 Hz using the {\it powspec} FTOOL task. 
To calculate the PDS we combined counts from the whole PCA energy range
using high resolution data modes.
PDS are normalized to units of fractional squared root-mean-square (RMS) per Hz.
We apply deadtime correction to PDS spectra according to \citet{zha95}.
Total RMS values were then  calculated as the square root of the integrated power in 
the frequency range from 0.01 Hz to 100 Hz. To investigate the energy distribution 
of the RMS (RMS spectrum) we first calculated spectra from the event or binned PCA modes and 
normalized the spectral bins with the RMS values calculated for each bin, thus obtaining 
the spectrum of the variable part of the emission. The RMS spectrum is then obtained as a ratio of the variable spectrum to the spectral model  for the total spectrum, obtained as described in the previous paragraph.

\section{Evolution of  XTE J1752-223}
\label{evolution}

In Figure \ref{curves} we present  {\it RXTE}/PCA and {\it Swift}/BAT lightcurves in various energy bands
as well as PCA hardness. %scenario for a BH outburst is a quick rise from quiescence  and immediate fast transition toHSS.
%The evolutionary pattern of the XTE J1752-223
%is fully consistent with the the behavior expected from the BH X-ray binary. Namely, 
The outburst
started in the LHS, which lasted from MJD 55130 to approximately MJD 55200.
% About  83\% of the
%the spectrum was non-thermal with the power law index of 1.4 and the cutoff energy of about 180 keV.
The PDS was represented by a  flat-top broken power law shape with a break at 0.18 Hz and 
total root-mean-square variability of 45\%.
%It is worth noting a long LHS exhibited by the source. 
%For technical reasons
%{\it RXTE} pointed at XTE J1752-223 for almost two days, making the longest uninterrupted 
%observation of the extreme LHS available to the astronomical community. 
%Study of timing properties for different energy ranges by \citet{md10} have 
%shown that the PDS shape in the LHS is energy independent and that the variability for the higher 
%energies lags softer emission in the manner similar to Cyg X-1. 
% Another famous BH 
%transient which exhibits such a long initial LHSs is GX 339-4 \citep{bel05b,bel06}. Similar 
%LHS exhibiting strong aperiodic variability has also been observed in Cyg X-1\citep{now99}. 
%As shown by the results of spectral modeling,  all parameters during the LHS stage
%were stable.

As indicated by the {\it Swift}/BAT lightcurve, beginning on MJD 55155  the source started to
evolve towards the IS \citep[see also][for detailed analysis of {\it Swift} data]{cur10}. 
{\it RXTE} observations were interrupted by Sun constraints until MJD 55215.
More coverage was provided by {\it Swift}/BAT monitoring, which was also not able to
observe from MJD 55175 to MJD 55194. For this reasons, it is hard to identify precisely 
the time when the source entered the IS.  We can only speculate that it most probably occurred
around MJD 55200, when a turnover in 15-200 keV {\it Swift}/BAT flux is observed.

RXTE monitoring resumed on MJD 55215.9 (19 Jan 2010 21:36 UT), when the source
was found to be in the fast transition phase. The first  {\it RXTE} pointed observation after interruption 
revealed a spectrum with a much softer power law component ($\Gamma=1.8$) and more 
pronounced disk black body part, which is consistent with the hard IS (HIMS) spectral state. 
The aperiodic  PDS component had a break at 0.5 Hz
and rms variability of 25\%. It was accompanied by a clear Type-C QPO \citep{cas05} at 2.2 Hz \citep{atel2391}.
Type-C QPOs were also observed during two following observations on MJD 55216.9 
(20 Jan 2010 21:36 UT) and 55217.9 (21 Jan 2010 21:30 UT)
 at 4.1 Hz and 5.3 Hz correspondingly, while the total rms variability decreased from 25\% to 18\%.
Near the end of the IS on MJD 55218 and 55220
Type A/B QPO were observed during observations 95360-01-01-00 and 95360-01-01-02.
These observations are consistent with a soft IS (SIMS).
After that the source entered the HSS. Some broad-band variability was observed at the level
of several percent through the first half of the HSS until MJD 55245. After that and until
the start of the decay stage IS the PDS from XTE J1752-223 was featureless
and consistent with zero variability above 0.01 Hz. It is interesting to note that the first part of the HSS when
the variability was present corresponds to the period of significant hard X-ray flux in the
energy range excluding the contribution from the thermal disk-black body component, i.e.
9-20 keV range for {\it RXTE}/PCA and 15-50 keV for {\it Swift}/BAT  (see Figure \ref{curves}). 
During the second non-variable part of the HSS the hard fluxes shown by both {\it RXTE} 
and {\it Swift} are close to the background level. Presence of significant  non-thermal emission during
the first half of the HSS is also indicated by the Comptonization factor $f$ in Figure \ref{fits}. Namely,
during the first "noisy" part of the HSS $f$ is between 3 and 6 percent, the second "quiet" half of the HSS is Comptonized
at $\sim$ 2\% only.

The source X-ray flux showed a gradual decline throughout the HSS,
as indicated by count rates in different channel ranges and the spectral model 
 normalization. On MJD 55282.56 (27 March 2010 13:25 UT), after more than 100 days in the HSS, the system
exhibited a sudden increase in flux, both in soft and hard energy ranges, such that the source
hardness increased. In addition, the aperiodic variability reappeared at the level of 8\% rms, indicating that the system was in the IS \citep{atel2518}. The decay stage IS lasted for approximately two weeks during which the Comptonized fraction
$f$ increased from below 0.02 to $\sim$0.7 and the power law index decreased from 2.0 to
1.6. The LHS was reached around April 4,  2010 (MJD 55295). The variability increased to 0.2-0.25 rms
represented by the flat-top noise with a hint of a broad QPO. 

\begin{figure}
\includegraphics[scale=0.33,angle=-90]{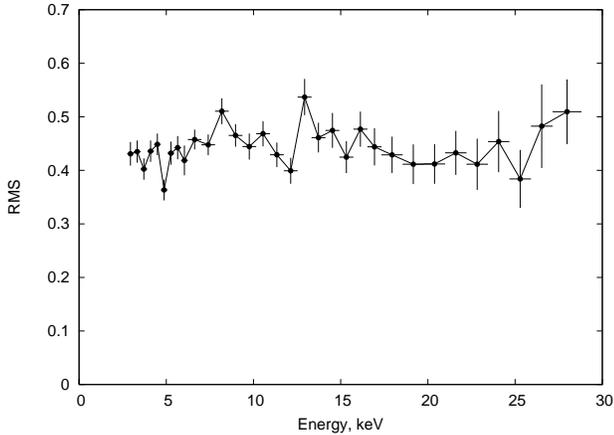}
\caption{ RMS spectrum in XTE J1752-223 during the LHS (Observation 94331-01-05-00).}
\label{rms_lhs}
\end{figure}

At the top of Figure \ref{corr} we show representative energy spectra and the PDS for each 
stage of the outburst. Each data point is shown in color according to its spectral state, i.e.
blue for the initial LHS, green for the first IS, red for the HSS, magenta for the second IS and
cyan for the decay LHS. The bottom diagrams show three correlation plots between essential
spectral and variability characteristics. One the left panel we present the spectrum 
normalization $N$ versus Comptonized fraction $f$ on the log-log scale.
On the middle diagram the $N-$rms correlation is shown and the right panel
presents the  rms-$f$ correlation.  As can be seen from the form of the correlation pattern,  the $N$-$f$ correlation is a Comptonization model analog of the famous q-shaped hardness-intensity diagram \citep[see][and references therein]{bel10}.
Both $N-f$ and $N-$RMS correlations show a strong hysteresis effect when the initial hard-to-soft spectral state transition occurs
at higher luminosity than the decay soft-to-hard transition \citep{m-h05}. The hysteresis effect is also 
seen in  the $N-$RMS diagram but is much less pronounced in the RMS-$f$
plot.  We discuss the evolution of the source through different spectral states in \S \ref{discussion}.

\section{XTE J1752-223 RMS spectrum}
\label{rms_dist}

\begin{figure}
\includegraphics[scale=0.33,angle=-90]{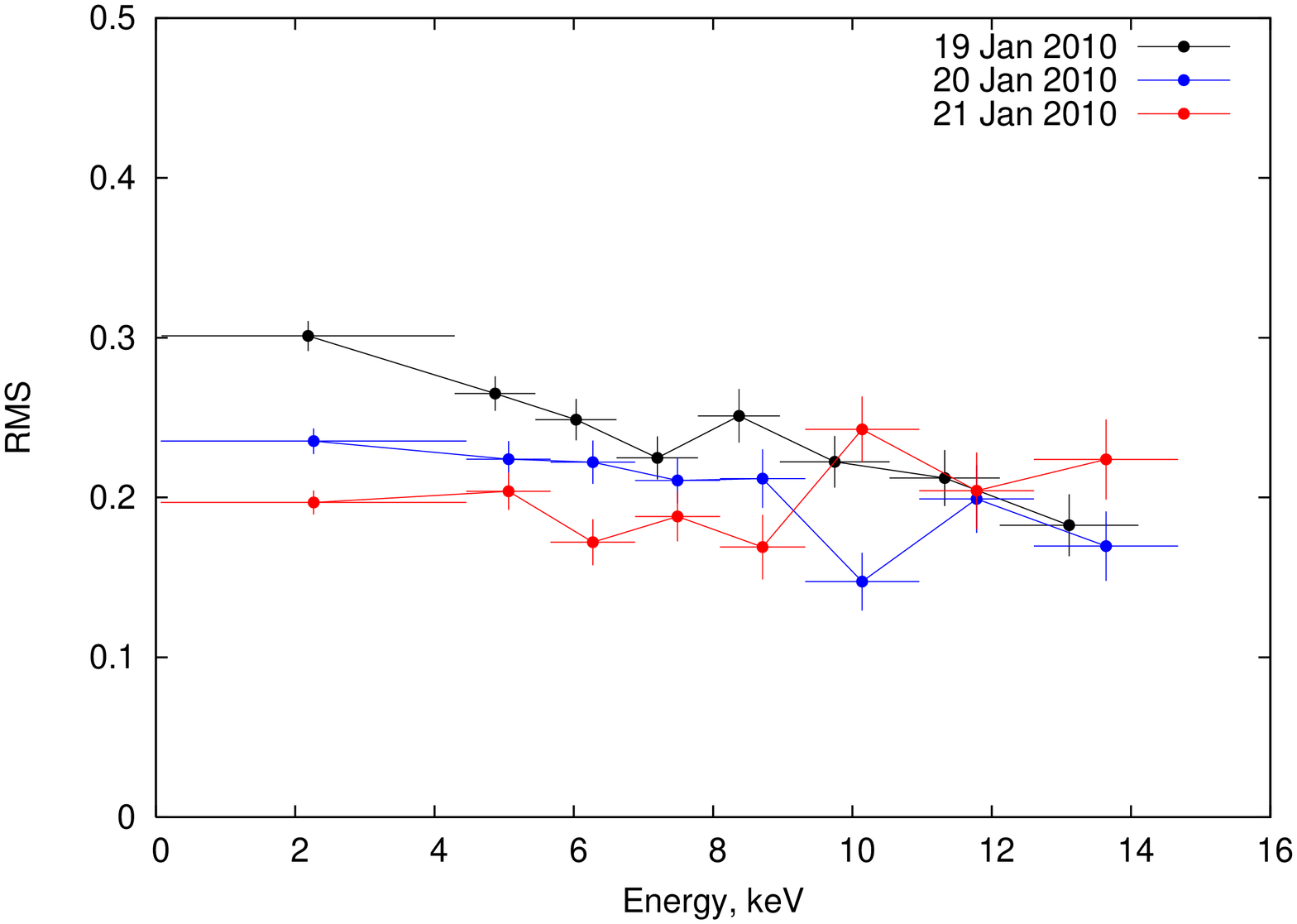}
\includegraphics[scale=0.33,angle=-90]{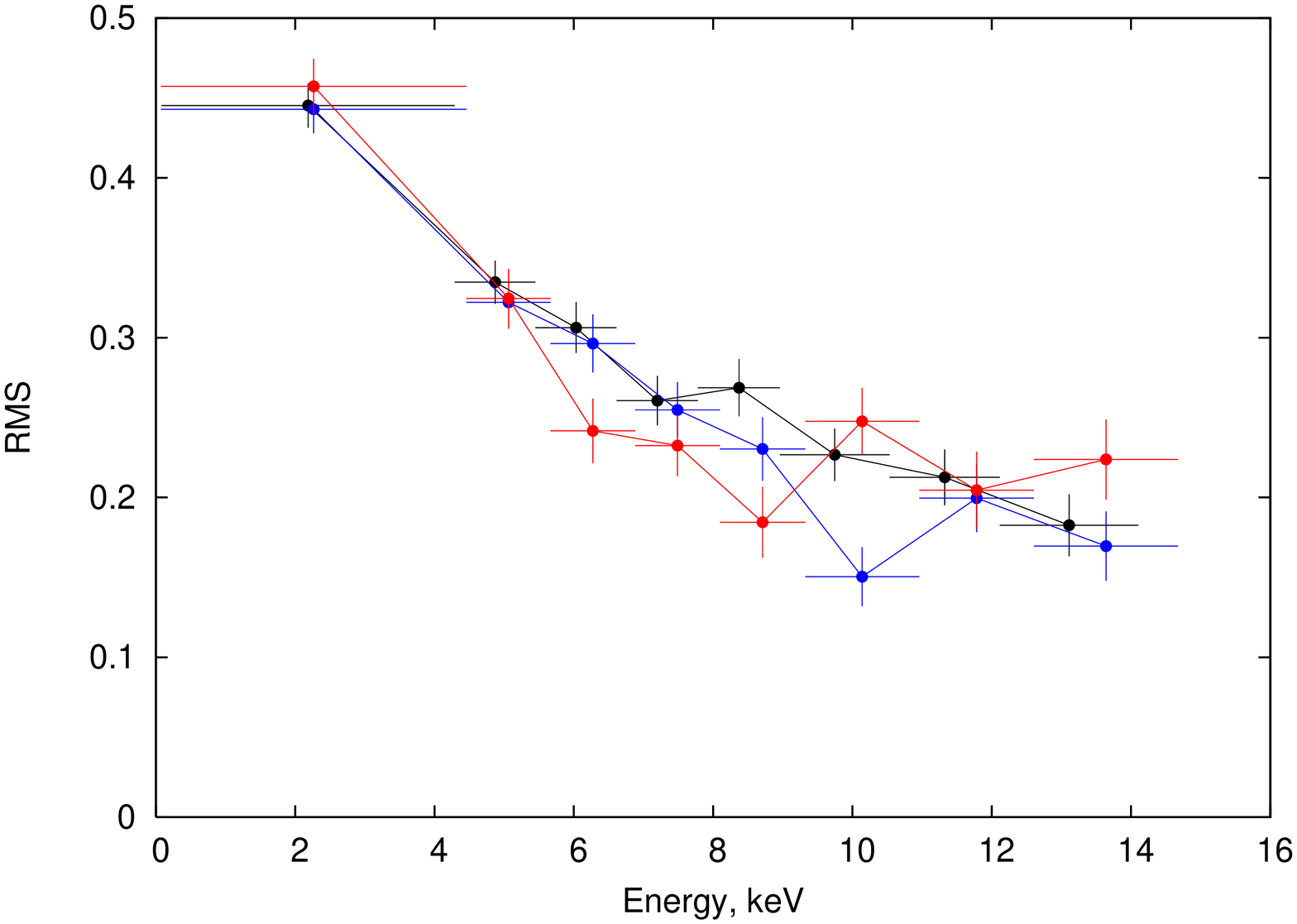}
\caption{ {\it Top:} Distributions of the fractional root-mean-square variability with energy for three 
observations during the spectral transition when  a Type-C QPO was observed (the hard IS).
The RMS is normalized by the total spectral model including both thermal and power law components (see text).
{\it Bottom:} The same as above but for RMS normalized by the power law component only.}
\label{rms_is}
\end{figure}

As noted by \citet{md10}, analysis of the {\it RXTE} data during the beginning of LHS stage in XTE J1752-223 
 showed no energy dependence of the variability versus energy.  We note, however, that according to analysis 
 of {\it Swift} data by  \citet{cur10}, the uniformity in RMS amplitudes versus energy
 breaks down for energies below 1.5 keV as the variability shown by XRT data is above 50\% during the LHS.
n Figure \ref{rms_lhs} we show the rms spectrum for the last {\it RXTE} observation 
during the LHS stage made on before the source became unobservable due to Sun constraint. 
Despite some scatter in RMS values the variability is practically constant in energy, at the level
of $\sim$0.45 rms.  However, the situation changes considerably during the IS. In Figure
\ref{rms_is} we show the rms energy distribution for the first three observations in the IS, when
Type-C QPOs were observed. The top panel shows the fractional RMS values calculated
in the conventional sense, i.e. the rms is normalized by a total source spectrum. It is clearly seen
that  rms values  for energies below 10 keV are gradually decreasing from 0.3 to 0.2 RMS, while the variability level
for the higher energies is stable at the level of $\sim$0.2. This decrease in variability is closely
correlated with the corresponding increase in contribution of the thermal "black-body" spectral
component in the lower energy range. This suggests that the intrinsic variability of the
thermal component is much less than than that in the power law component. This is consistent
with the previous claims of stability of the accretion disk emission \citep{chur01,hom01,zyc07}. 

Among the observational facts  which indicate that
fast variability in XTE J1752-223 is concentrated towards (if not entirely confined within) the non-thermal 
part of the spectrum are  (a) very high variability in the LHS, (b) non-detection of the variability
in the second half of the HSS, when the Comptonization factor $f$ is less than 3\%, and (c)
tight correlations between $f$ and rms (Figure \ref{corr}, right panel in the bottom).
It is therefore more informative and natural to consider the rms variability level with respect to the power law 
emission component only. The rms spectra calculated with respect to the non-thermal spectral 
component (i.e. with the black body and Gaussian components excluded) are shown 
in the bottom panel of Figure \ref{rms_is}. We first note the remarkable similarity 
of the power-law normalized rms spectra for all three data sets. This fact shows that,
if the variability is indeed concentrated in the power law component, the rms energy distribution
remains fairly constant during the considered part of the IS. The second notable result 
is that the highest level of variability is observed for the lowest energy $\sim 3$ keV and is the same
that the variability observed during the LHS, i.e. about 0.45. The third fact to mention is the gradual
decrease in rms with energy. These facts have strong implications for the scenarios 
of the non-thermal emission production in BH X-ray binaries, as discuss  in \S \ref{discussion}.

\section{BH mass and the distance to XTE J1752-223}
\label{mass} 

BH mass and distance measurements from X-ray spectral 
and fast variability analysis data require the following data and information.
First, we need a  reference source  for which the BH mass and distance are known.
For the reference and the target (the source for which the fundamental parameters
are to be determined) we need scalable and well sampled observations of timing and spectral evolution through
spectral transitions. The scaling factors are calculated between the spectral index - QPO frequency and index - normalization
correlation patterns, which is achieved usually by fitting an empirical analytical function to 
the data. While the QPO frequency depends on the BH mass ratio only, the normalization ratio is a function
of the masses and distances and also contains a factor depending on inclination
angles. This geometrical factor is very close to unity in most cases (see ST09 for details).

\begin{figure}
\includegraphics[scale=0.33,angle=-90]{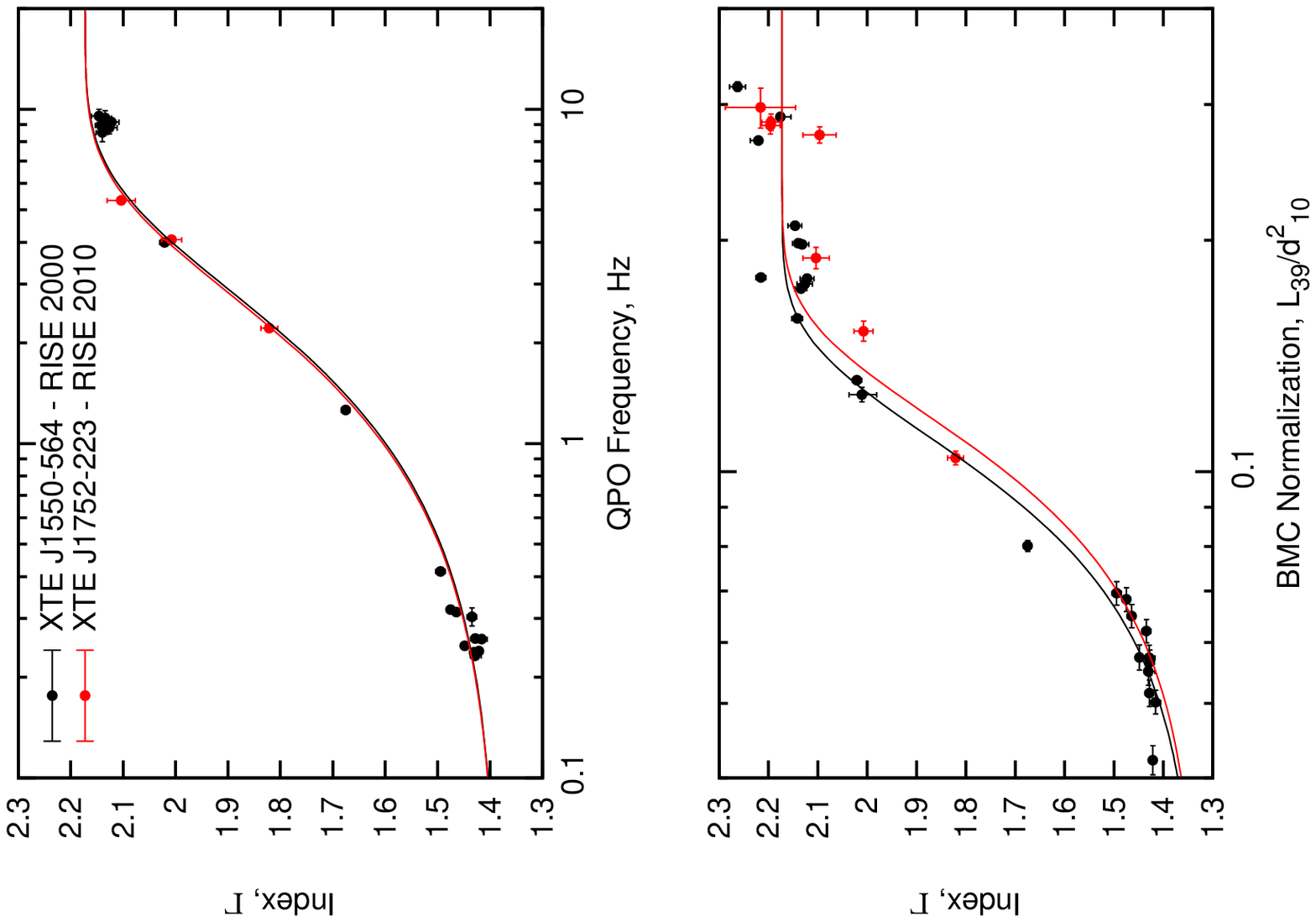}
\includegraphics[scale=0.33,angle=-90]{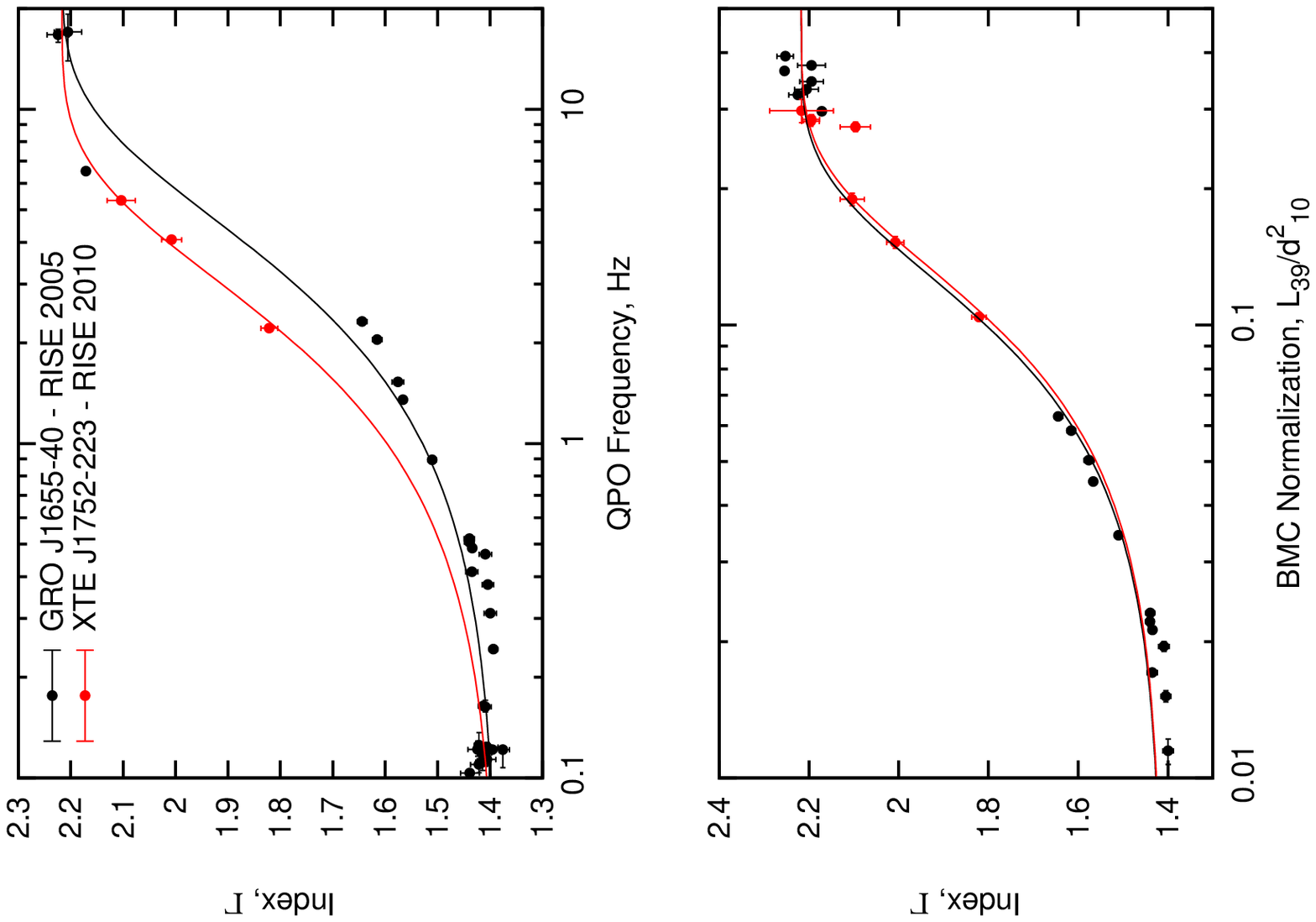}
\caption{Correlation scaling of the state transition data for   XTE J1752-223, presented by black points, versus GRO J1655-40 (left panel) and  XTE J1550-564 (right panel). Data for reference sources are shown in red.  Top panels show scaling  index-QPO frequency curves, while at the bottom we show Index vs spectrum normalization data.}
\label{scaling}
\end{figure}

As mentioned above, {\it RXTE} was not able to observe an entire hard-to-soft transition episode.
QPOs were detected for three observations only. These data,
however, provide enough information to test scalability and calculate scale coefficients
in the index-QPO domain. The index-normalization data allow more reliable examination
of the patterns self-similarity. To apply the scaling method for BH mass and distance measurement in XTE J1752-233 we
use data from LHS-HSS transitions in two BH candidates GRO J1655-40 (2005 outburst) and 
XTE J1550-564 (2000 outburst) for which independent 
information on the BH mass and distance is available. These transitions were
analyzed in ST09.  Both data sets show correlation patterns scalable to the XTE J1752-223 transition behavior. 
We fit the reference correlation patterns with the analytical function discussed in ST09
and then calculate the scaling coefficient by refitting the functional shape to the target source data using 
transform $x\rightarrow s x$, where $x$ is a x-axis argument and $s$ is a corresponding scaling coefficient. 
The data  and the approximating functions used for scaling are shown in Figure \ref{scaling}.
The resulting values of the scaling coefficients and the final values of BH mass and distance for XTE J1752-223 are given
in Table \ref{table1}. The BH mass and distance values are calculated and errors for all parameters are propagated according
to formulas $M_t=s_\nu M_s$ and $d_t=(s_\nu s_n)^{1/2}d_r$ (assuming the unity geometrical factor $f_G$, see ST09),
where the subscript $t$ stands for the target source (i.e. XTE J1752-223)
and $r$ is for the reference source (i.e. GRO J1655-40 or XTE J1550-564).
All errors represent 1$\sigma$ confidence parameter ranges. 
Both scalings  based on the independently determined reference values give
markedly consistent results for both BH mass and the source distance. We
conclude that the BH mass in XTE J1753-223 is $M_{XTE J1752-223} = 9.6 \pm 0.9 M_{\odot}$,
while the distance is $d_{XTE J1752-223} = 3.5 \pm 0.4$ kpc, which are obtained by averaging the results
of scaling to GRO J1655-40 and XTE J1550-564. 

The agreement of our double validation for the obtained BH  mass and distance is rather convincing. 
We note, however, that the hard-to-soft transition that we use for our calculations 
was not fully sampled due the Sun constraints. We also note that for the 2005 transition in GRO J1655-40
one data point falls off the main $\nu_{QPO}$-index correlation  track. This observation ({\it RXTE} ObsID 91702-01-02-00G)
was performed during  a quick flare, when the source was undergoing fast changes in its spectral and variability
properties \citep[see][]{sh07}. Therefore, the measurements of QPO frequency and spectral index may be biased
during this observation.
%While the QPO during this 
%observation was accompanied by strong noise and was quite broad.  It is not clear therefore if
%this QPO feature was of the Type-C \citep{cas05}. 
Nevertheless, we include the data point for this observation in the 
scaling procedure for consistency with previous analyses.
%X-ray flux exhibited a flux increase by a factor of two. While the spectral index also increased
%the QPO did not show the expected rise in frequency while became broad instead.
%This is reminiscent of Type-B QPO behavior. 
%The QPO frequency-index  datum 
%does falls slightly off the main correlation track onto 
Optical monitoring holds promise for dynamical mass 
determination \citep{atel63}. Pending such a confirmation, some caution is recommended when
interpreting the above results.

\begin{table}
\tablewidth{0pt}
\caption{BH Mass and Distance Measurement in XTE J1752-223}
\begin{tabular}{lll}
\hline
Reference Source & Scaling  & XTE J1752-223 \\
BH mass and Distance$^a$ & Coefficients & BH mass and distance $^b$\\ 
\hline
GRO J1655-40 &&\\
\hline
$M_{1655} = 6.3\pm0.5$\ (1)  & $s_\nu = 1.50 \pm 0.04$& $M_{1752}=9.4\pm1.0$ \\
$d_{1655} = 3.2\pm 0.2$\ (2) & $s_N = 0.96 \pm 0.05$ & $d_{1752}=3.8\pm0.4$ \\
\hline
XTE J550-564&&\\
\hline
$M_{1550} = 9.5\pm1.1$\ (3) & $s_\nu = 1.03 \pm 0.03$ & $M_{1752} = 9.8\pm1.4$ \\
$M_{1550} = 3.3\pm0.5$\ (4) & $s_N = 0.94 \pm 0.05$ & $ d_{1752} = 3.2\pm0.7$  \\
 \hline\\
 \end{tabular}
 
 $^a$ $m = M/M_\odot$, distance $d$ is given in kiloparsecs 
 
References - $^1$ \citet{gre01}, $^2$ \citet{hje95}, $^3$ \citet{oro02}, $^4$ ST09
\label{table1}
\end{table}

\section{Discussion} 
\label{discussion}

Let us summarize the main results of our observations and analysis.
During the reported discovery outburst a new galactic BH candidate source
XTE J1752-223 has shown behavior completely
consistent with the general pattern of previously observed outbursts from BH X-ray binaries.
It has evolved through BH spectral states, showing the expected corresponding
changes in variability. 

The characteristic extreme LHS was observed in the beginning 
of the outburst. This state was seen in many galactic  BH  transients during the outburst initial phase.
In most cases the source evolves quickly towards softer states.  However, the 
LHS in XTE J1752-223 lasted for 3 months, showing uniquely stable spectral and timing 
properties \citep[see also][]{md10}. Prolonged hard states have been
seen in GX 339-4 \citep{bel06}, XTE J1118+480 \citep{bro10} and Cygnus X-1 \citep{now99}. 
XTE J1118+480 never showed a transition to HSS while Cygnus X-1 is a persistent
source. Therefore, the closest analogy is with GX 339-4. The duration of the LHS phase may  be related to the time delay 
between the initial mass injection by a donor star into the Roche lobe and the time for the thin accretion disk
to form and to reach the inner accretion flow zone and to trigger a state transition.
The gas with high angular momentum has to form a thin disk in order to dispose excessive momentum
before being accreted. On the other hand, accretion of the low momentum gas occurs in the non-keplerian
regime and the matter reaches the compact object much faster. 

The emission during the LHS is quite variable with a fractional RMS of 43-48\%.
In Figure \ref{rms_lhs} we show the RMS spectrum during the last LHS observation 
on November 20 2:24 UT (ObsID 94331-01-05-00). The distribution is uniform
over the energy range from 3 to 30 keV. However, during the state transition 
the rms spectrum changes considerably. As the source evolves towards the HSS, the overall RMS level drops.
At the later stages of the IS, which were observed by {\it RXTE}, the drop in the variability amplitude
is consistent with the increase in the soft thermal component due to an accretion disk as shown
in \S \ref{rms_dist}. In fact, we find evidence that the entire variability is contained
within the power law spectral component. When we normalize the RMS spectrum  
by the non-thermal part of the spectrum the RMS energy distribution is non-uniform with the RMS amplitude
decreasing with energy. This fact may be related to the physics of photon upscattering.
\citet{motta09} pointed out the evolution pattern of the high energy
cutoff of the energy spectrum during the state transition shown by GX 339-4 during the 2007
outburst. The authors observed a decrease in the cutoff energy during the first
part of the transition (hard IS) from about 150 keV to 50 keV and then an abrupt cutoff increase 
during the second part of the transition (soft IS), which was also accompanied by
changes in timing behavior. \citet{ts10} have recently identified the same effect in the
galactic BHC XTE J1550-564 and interpreted it as a manifestation of a gradual change in  
the photon upscattering  regime from pure thermal to bulk motion  Comptonization \citep{bmc}.
The evolution of the rms spectrum shown by XTE J1752-223 during the hard-to-soft
state transition can also be interpreted in terms of this scenario. During the LHS 
Comptonization is purely thermal. Photons detected at all energies are upscattered 
by the same hot turbulent plasma resulting in a uniform distribution in rms amplitudes versus energy (see Figure \ref{rms_lhs}).
As the source moves towards the softer states,  emission from an 
advancing accretion disk effectively cools the Compton corona causing its contraction and exposing
the innermost region of the accretion flow to the observer, where bulk motion is the dominant process for photon upscattering. Therefore, the gradual collapse of the thermal corona leads to the increasing contribution of the bulk motion Comptonization as
an upscattering agent for the emerging non-thermal radiation. The bulk motion contribution in upscattering should
also be higher for the photons emerging at higher energies. 
If we assume that the variability is mostly produced by perturbations
in the thermal Comptonization region and the bulk motion region is much more stable, the observed 
RMS decline with energy seen in Figure \ref{rms_is} finds a natural explanation. Namely,
the higher energy emission is less variable as the photons are boosted in the less variable bulk motion 
region.  It is also clear that the soft photons at about 3 keV which belong to the power spectral component
are produced by the thermal Comptonization as revealed by the corresponding variability of $\sim$45\%
consistent with RMS level of the LHS. Further investigation of the RMS spectra evolution along these lines
seems to be very promising. We will pursue this to verify the RMS behavior observed in XTE J1752-223 in
other galactic  BH sources.

\section{Conclusions} 
\label{summary}

XTE J1752-223 is a new galactic BH candidate. During the 2009-2010 discovery outburst the source
evolved from the extreme LHS to HSS through IS as observed by {\it RXTE} extensive monitoring program.
We studied the evolution of spectral and timing properties of the source during the 
the outburst  using available {\it RXTE} data, specifically concentrating on the transition phase, when the source 
exhibited low-frequency QPOs. Evolution of the RMS variability throughout the outburst  and the behavior of the RMS 
spectrum during the state transition suggest the non-thermal part of the energy spectrum as a primary source of the flux modulation.
QPO frequency and spectrum normalization evolution, when
compared to the transition data from 2005 outburst of GRO J1655-40 and 
2000 outburst of XTE J1550-564, strongly indicates that BH mass in XTE J1752-223 is $9.6\pm0.9$ solar masses.
We estimate the distance to the source in the range of $3.5\pm0.4$ kpc. This result confirms the expectation that the
compact object in XTE J1752-223 is a stellar mass BH.

We made use of the data provided through HEASARC and {\it RXTE} SOF. NS acknowledge the support of
this work  by NASA grant NNX09AF02G.

%\newpage

%\newpage

%\newpage

%\newpage

%\newpage

\end{document}